\documentclass[12pt]{article}
 
\usepackage{graphicx}
\usepackage{amsfonts}
\usepackage{amsthm}
\usepackage[top=3cm, bottom=3.25cm, left=2.75cm, right=2.75cm]{geometry} 

\newcommand{\Section}[1]{\section{#1} \setcounter{equation}{0}} 
\newcommand{\beq}{\begin{equation}} 
\newcommand{\eeq}[1]{\label{#1}\end{equation}}
\newcommand{\ber}{\begin{eqnarray}} 
\newcommand{\eer}[1]{\label{#1}\end{eqnarray}}

\newcommand{\ft}[2]{{\textstyle\frac{#1}{#2}}}
\newcommand{\nll}{N\!=\!(1,1)}
\newcommand{\nZZ}{N\!=\!(2,2)}
\newcommand{\vf}{\varphi}
\newcommand{\ff}{\mathbb{F}}
\newcommand{\ffb}{{\bar\mathbb{F}}}
\newcommand{\fft}{{\tilde\mathbb{F}}}
\newcommand{\fftb}{{\bar{\tilde\mathbb{F}}}}

\newcommand{\vv}[1]{\mathbb{V}^{#1}}
\newcommand{\vvb}[1]{\bar{\mathbb{V}}^{#1}}
\newcommand{\vvt}[1]{\tilde{\mathbb{V}}^{#1}}
\newcommand{\vvtb}[1]{\bar{\tilde{\mathbb{V}}}^{#1}}

\newcommand{\dt}[1]{\hat{d}^{#1}}
\newcommand{\dq}[1]{\hat{q}^{#1}}
\newcommand{\jj}[3]{{J}_{#1 #3}^{#2}}
\newcommand{\bbD}[1]{\mathbb{D}_{#1}}
\newcommand{\bbDB}[1]{\bar{\mathbb{D}}_{#1}}
\newcommand{\bbG}[1]{\mathbb{G}_{#1}}
\newcommand{\bbGB}[1]{\bar{\mathbb{G}}_{#1}}
\newcommand{\bbX}[1]{\mathbb{X}_{#1}}
\newcommand{\bbXB}[1]{\bar{\mathbb{X}}_{#1}}
\newcommand{\bbXh}[1]{\tilde{\mathbb{X}}_{#1}}
\newcommand{\bbXBh}[1]{\bar{\tilde{\mathbb{X}}}_{#1}}

\newcommand{\hee}[1]{\Xi^1_{#1}}
\newcommand{\hcc}[1]{\Xi^2_{#1}}

\newcommand{\hcd}[1]{\nabla_{#1}}
\newcommand{\hPh}[1]{\varphi^{#1}}

\def\+{{+\!\!\!+}} 
 
\newcommand{\aleq}{&\!\!\!=\!\!\!&}
\newcommand{\nn}{\nonumber}
\newcommand{\kah}{K\"ahler~}
\newcommand{\pa}[1]{\partial_{#1}}
\newcommand{\lam}{\Lambda}
\newcommand{\G}{\Gamma}
\newcommand{\lamb}{\bar\Lambda}
\newcommand{\lamt}{\tilde\Lambda}
\newcommand{\lamtb}{\bar{\tilde{\Lambda}}}

\newcommand{\eg}{\textit{e.g.},~}

\begin{document}
\renewcommand{\theequation}{\thesection.\arabic{equation}}  
\setcounter{page}{0}
\thispagestyle{empty}

\begin{flushright} \small
UUITP-12/07 \\NORDITA-2007-22\\  YITP-SB-07-22\\ 
\end{flushright}
\smallskip
\begin{center} \LARGE
{\bf  T-duality and Generalized \kah Geometry}
 \\[12mm] \normalsize
{\bf Ulf~Lindstr\"om$^{a,b,c}$, Martin Ro\v cek$^{d}$, Itai Ryb$^{d}$,\\
Rikard von Unge$^{e}$, and Maxim Zabzine$^{a}$} \\[8mm]
{\small\it
$^a$Department of Theoretical Physics 
Uppsala University, \\ Box 803, SE-751 08 Uppsala, Sweden \\
~\\
$^b$HIP-Helsinki Institute of Physics, University of Helsinki,\\
P.O. Box 64 FIN-00014  Suomi-Finland\\
~\\
$^c$NORDITA, Roslagstullsbacken 23,\\
SE-10691 Stockholm, Sweden\\
~\\
$^d$C.N.Yang Institute for Theoretical Physics, Stony Brook University, \\
Stony Brook, NY 11794-3840,USA\\
~\\
$^{e}$Institute for Theoretical Physics, Masaryk University, \\ 
61137 Brno, Czech Republic \\~\\}
\end{center}
\vspace{10mm}
\centerline{\bf\large Abstract} 
\bigskip
\noindent  
We use the new $\nZZ$ vector multiplets to clarify T-dualities for 
generalized \kah  geometries. Following the usual procedure, we gauge isometries 
of nonlinear $\sigma$-models and 
introduce Lagrange multipliers that constrain the field-strengths of the gauge fields 
to vanish. Integrating out the Lagrange multipliers leads to the original action, 
whereas integrating out the 
vector multiplets gives the dual action. 
The description is given both in $\nZZ$ and $\nll$ superspace.
\eject
\normalsize

%\addtocontents{toc}
%\ofcontents

%\end{titlepage}

\eject

\section{Introduction}

The basic inspiration for our work is the interesting duality found in 
\cite{Grisaru:1997ep,Bogaerts:1999jc} for two dimensional nonlinear $\sigma$-models
with $\nZZ$ supersymmetry and target space
geometries that are not K\"ahler. As was shown in \cite{hklr,Rocek:1991ps},
T-dualities arise when one gauges an isometry, and then constrains
the field-strength of the corresponding gauge field to vanish.  
In this paper, we use the new vector multiplets introduced in \cite{Lindstrom:2007vc,Gates:2007ve} to describe T-duality for 
generalized \kah geometries (for a sampling of articles in the field, see \cite{genk}). 
We first work in $\nZZ$ superspace, and then reduce
to $\nll$ superspace and find the usual T-duality of Buscher \cite{Buscher}. 

The plan of the paper is as follows: In the next section we briefly review T-duality 
in the pure \kah case \cite{Buscher}. We then review the classes of 
isometries that generalized \kah geometries admit. Next, we consider T-dualities 
along isometries in the kernel of the commutator of the left and right 
complex structures that mix chiral and twisted chiral multiplets \cite{ggg}. Finally 
we describe T-dualities along isometries in the cokernel of the commutator, 
which act only on the semichiral multiplets\cite{bbb}. We use the bihermitian
description of generalized K\"ahler geometry throughout the paper, and leave the description
of T-duality in terms of generalized complex structures to the future.

We end with a brief conclusion. 

\Section{\kah geometry and T-duality}
In this section, we briefly review isometries, gauging, and T-duality in $\nZZ$ and $\nll$ superspace for a system with chiral superfields $\phi^a$ and
an $\nZZ$ superspace Lagrange density given by a 
\kah potential $K(\phi^a,\bar{\phi}^a)$ \cite{Hull:1985pq,Rocek:1991ps,Buscher}.
For simplicity, we consider an isometry generated by a holomorphic Killing vector $k$ that leaves the \kah potential invariant\footnote{The general case when
$K$ is invariant only up to a \kah transformation is discussed in detail in \cite{Hull:1985pq},
and in the generalized \kah case in \cite{LRRvUZ}. }
\beq
k\equiv k^i \pa{i} = k^a \pa{a} + \bar k^a \bar\pa{a}
~~,~~~ \mathcal{L}_k K = 0 ~,
\eeq{eq::isometry}
where $\vf^i=\{\phi^a,\bar\phi^a\}$.
The isometry is gauged using a multiplet $V^\phi$ to promote the constant
(real) transformation parameter
$\lambda$ to a complex chiral superfields $\lam$:
\beq
\lambda(k^a \pa{a} + \bar k^a \bar\pa{a}) K(\phi^a,\bar{\phi}^a)=0~~\to~~
\left(\lam k^a\pa{a}+\lamb \bar k^{a}\bar\pa{a} + \delta V^\phi \pa{V^\phi}\right)\! 
K^{(g)}(\phi^a,\bar{\phi}^a,V^\phi) = 0 ~.
\eeq{eq::gauged_kahler_equations}
From (\ref{eq::isometry}), it follows 
that\footnote{If $K$ is not invariant, in (\ref{kfid}) $\mathcal{L}_{Jk}K$
must be replaced by the moment map of $k$:
$\mathcal{L}_{Jk}K\to-\mu$.}
\beq
\left(\lam k^a\pa{a}+\lamb \bar k^{a}\bar\pa{a}
\right) K(\phi^a,\bar{\phi}^a)=\ft{i}2(\lamb-\lam)\mathcal{L}_{Jk} K~~,~~~
Jk=i(k^a\pa{a} - \bar k^{a}\bar\pa{a} )~.
\eeq{kfid}
Using the usual gauge transformation $\delta V^\phi = i(\lamb-\lam)$, we find the gauged
action \cite{Hull:1985pq}:
\beq
K^{(g)}(\phi^a,\bar{\phi}^a,V^\phi) = \exp \left(- \ft12 V \mathcal{L}_{Jk} \right) 
K(\phi^a,\bar{\phi}^a) ~.
\eeq{eq::gauged_action_and_transformation}
To find the T-dual model \cite{Rocek:1991ps}, we constrain the twisted chiral
field-strength $\bbDB+\bbD-V^\phi$ to vanish. We impose this with a Legandre transformation  of the density with a twisted chiral Lagrange multiplier $\chi$:
\begin{equation}
\label{eq:constrained_lagrangian}
K^{(g)}(\phi^a,\bar{\phi}^a,V^\phi) - (\chi+\bar{\chi})V^\phi~.
\end{equation}
In $\nZZ$ superspace, we find the T-dual Lagrange densities by integrating 
out either $\chi+\bar{\chi}$,
which gives the original \kah potential, or $V^\phi$, which gives the 
T-dual potential $\tilde{K}(\chi+\bar{\chi},x^A)$
where $x^A$ are ``spectator'' fields, {\it i.e.}, 
combinations of the $\hPh{i}$ that are inert under the
action of the isometry (\ref{eq::isometry}). The geometric 
nature of the duality is made manifest when we descend to $\nll$ 
superspace. In Wess-Zumino gauge, the
$\nll$ components of the multiplet $V^\phi$ and the covariant derivatives are
\begin{equation}
V^\phi | = 0 ~~,~~~ Q_\pm V^\phi | = A_\pm ~~,~~~ 
i\, Q_+ Q_- V^\phi | = d ~~,~~~ 
\hcd{\pm}\hPh{i} = D_\pm \hPh{i} - A_\pm k^i 
\end{equation}
the constrained Lagrange density (\ref{eq:constrained_lagrangian}) becomes:
\begin{equation}
\label{eq:pure_reduce}
g_{ij} \hcd{+}
\vf^i\hcd{-}\vf^j - i\,d (K_i \jj{}{i}{j} k^j + (\chi + \bar{\chi}))
+ f (\chi-\bar{\chi}) 
\end{equation} 
where $f=i(D_+A_-+D_-A_+)$ is the $\nll$ field-strength for the gauge fields, 
$g_{ij}$ is the \kah metric, and $K_i \jj{}{i}{j} k^j\equiv\mathcal{L}_{Jk}K$ is 
proportional to the moment map when the \kah potential is invariant (as discussed above,
in general $\mathcal{L}_{Jk}K\to-\mu$).  
Integrating out the $\nll$ auxiliary superfield $d$ sets 
$\chi+\bar{\chi}$ equal to the moment map. This can be solved either by expressing
$\chi+\bar{\chi}$ as a function of $\vf^i$, or by changing coordinates to $\chi+\bar{\chi}$
and a combination of $\vf^i$ algebraically independent of the moment map; the two procedures
are related simply by a diffeomorphism. 
This gives the $\nll$ gauged Lagrange density with Lagrange multiplier 
$(\chi-\bar{\chi})$ constraining the field-strength $f$:
\begin{equation}
\mathcal{L}_1 = g_{ij} \hcd{+}\vf^i\hcd{-}\vf^j + f (\chi-\bar{\chi})~,
\end{equation}
where $g_{ij}$ is the original K\"ahler metric (in $\vf^i$ coordinates).
Thus $\nZZ$ T-duality is the same as $\nll$ T-duality up to an accompanying diffeomorphism;
this was originally proven by Buscher \cite{Buscher}, but not explicitly spelled out.

\Section{T-duality for the generalized \kah geometry}
In a recent paper \cite{Lindstrom:2007vc} we discussed gauge multiplets suitable
for gauging isometries of generalized \kah geometries. We found three distinct
vector multiplets, corresponding to three distinct types of isometries: those along the
kernel of either $J_+-J_-$ or (equivalently) $J_++J_-$, those acting on both kernels,
and those along the cokernel of the commutator $[J_+,J_-]$. The isometries can be expressed, following \cite{Grisaru:1997ep}, in adapted coordinates:
\begin{eqnarray}
k_\phi\aleq i(\pa{\phi}-\pa{\bar\phi})~,\\[1mm]
k_{\phi \chi}\aleq i(\pa{\phi}-\pa{\bar\phi}-\pa{\chi}+\pa{\bar\chi}) ~,\\[1mm]
k_{LR}\aleq i(\pa{L}-\pa{\bar{L}}-\pa{R}+\pa{\bar{R}}) ~.
\end{eqnarray}
If we assume that the generalized \kah potential is invariant, the corresponding gauged
actions are:
\begin{eqnarray}
K_\phi \aleq K_\phi \left( \phi + \bar{\phi} + V^\phi,x\right)~,\label{kf}\\[1mm]
K_{\phi\chi} \aleq K_{\phi\chi} \left( \phi + \bar{\phi} + 
V^\phi,\chi + \bar{\chi}+ V^\chi, i(\phi - \bar{\phi} + 
\chi - \bar{\chi}) + V^\prime,x\right)~,\label{kfc}\\[1mm]
K_\mathbb{X} \aleq K_\mathbb{X} \left( \bbX{L}+\bbXB{L}+\vv{L} , 
\bbX{R}+\bbXB{R}+\vv{R} , 
i(\bbX{L}-\bbXB{L}+\bbX{R}-\bbXB{R})+\vv{\prime},x\right)~, \label{kx}
\end{eqnarray}
where $x$ represents all possible spectator fields.
The case (\ref{kf}) is essentially identical to the \kah case above; aside from 
subtleties pertaining to the interpretation of the moment map, 
which will be discussed in \cite{LRRvUZ}, there are no new features. We now
consider (\ref{kfc},\ref{kx}) in detail, and show that they again reduce to standard
Buscher duality in $\nll$ superspace, along with some natural diffeomorphisms inherited from
$\nZZ$ superspace.
A more general discussion of isometries and moment maps will be given in \cite{LRRvUZ}. 
\subsection{T-duality for an isometry $k_{\phi\chi}$}
For an invariant generalized \kah potential $K$ in adapted coordinates, the gauged action
is (\ref{kfc}). In the special circumstance when all the 
spectators are (twisted) chiral, we can give a nice geometric interpretation of the
gauging analogous to the \kah case above.
In this case both complex 
structures are simultaneously diagonalizable; 
and the manifold has the \textbf{Bi}hermitian 
\textbf{L}ocal \textbf{P}roduct (BiLP) 
geometry defined in \cite{Lindstrom:2007qf}.
Using the invariance of $K$ under $k_{\phi\chi}$, 
and using the complex structures $J_\pm$ and their product $\Pi = J_+ J_-$
\begin{eqnarray}
 \label{kfcK}
 [i(\lam \pa{\phi}\!\!\!&-&\!\!\!\lamt\pa{\chi})+c.c.]K\\
\aleq\!\!\!\ft{i}4[ (\lamb-\lam) 
 \mathcal{L}_{(J_+ + J_-)k} 
 + (\lamtb-\lamt) \mathcal{L}_{(J_+ -J_-)k}
 +i(\lam+\lamb-\lamt-\lamtb)\mathcal{L}_{\Pi k}] K\nn  
\end{eqnarray}
To gauge the isometry, we require 
\ber
&&\!\!\!\!\!\!0=\delta V^\alpha \pa{{V^\alpha}}K^{(g)}\\
&&~~+\ft{i}4[ (\lamb-\lam) 
 \mathcal{L}_{(J_+ + J_-)k} 
 + (\lamtb-\lamt) \mathcal{L}_{(J_+ -J_-)k}
 +i(\lam+\lamb-\lamt-\lamtb)\mathcal{L}_{ \Pi k}] K^{(g)}\nn
 \eer{kfcKg}
The three superfields of the large vector multiplet \cite{Lindstrom:2007vc} 
have the right gauge transformations to gauge this symmetry:\footnote{
Our conventions here, which are compatible with the inherent geometric 
objects $J_\pm , k$, are slightly different than those introduced in 
\cite{Lindstrom:2007vc}; see Appendix \ref{table} for the relation between the conventions.}
\begin{eqnarray}
 \delta V^\phi = i (\lamb-\lam) ~~,~~~ 
 \delta V^\chi = i(\lamtb-\lamt) ~~,~~~ 
 \delta V^\prime = (-\lam-\lamb+\lamt+\lamtb) \nn \\
\Rightarrow K^{(g)} = \exp \left( -\ft{1}{4}V^\phi 
\mathcal{L}_{(J_+ + J_-)k} -\ft{1}{4}V^\chi 
\mathcal{L}_{(J_+ - J_-)k} - \ft{1}{4} V^\prime 
\mathcal{L}_{\Pi k} \right)K~.
\end{eqnarray}

To find the T-dual, we introduce Lagrange multipliers 
that constrain the field strengths of the large vector multiplet to vanish. 
As discussed in \cite{Lindstrom:2007vc}, it is useful to introduce 
complex potentials for the field-strengths:
\begin{eqnarray}
V_L = \frac{1}{2} [-V^\prime +i(V^\phi-V^\chi)] 
&\Rightarrow& \delta_g V_L = \lam-\lamt~, \nn\\ 
V_R = \frac{1}{2} [-V^\prime +i(V^\phi+V^\chi)] &\Rightarrow& \delta_g 
V_R = \lam-\lamtb ~.
\end{eqnarray} Since $(\lamt)\lam $ are respectively (twisted)chiral, these give the following gauge invariant complex spinor, semichiral, field-strengths:
\begin{eqnarray}
\bbG{+} \aleq \bbDB{+} V_L ~~,~~~~ \bbGB{+} = \bbD{+} \bar{V}_L~,\nn\\
\bbG{-} \aleq \bbDB{-} V_R ~~,~~~~ \bbGB{-} = \bbD{-} \bar{V}_R~.
\end{eqnarray}
Using the chirality properties of the field-strengths we obtain the constrained $\nZZ$ generalized \kah potential, as in (\ref{eq:constrained_lagrangian}), using semichiral Lagrange multipliers $\bbXh{}$:
\begin{equation}
\label{eq:large_kahler_pot}
K^{(g)} - \mathcal{L}_{const.} = K^{(g)} - \ft{1}{2}\bbXh{L} V_L - \ft{1}{2}\bbXBh{L} \bar{V}_L - \ft{1}{2}\bbXh{R} V_R - \ft{1}{2}\bbXBh{R} \bar{V}_R~.
\end{equation}
This applies to the general case, not just BiLP geometries, though in general,
we do not have a nice geometric 
form of $K^{(g)}$ (this will be discussed in \cite{LRRvUZ}).
\subsubsection{Reduction to $\nll$ superspace}
Using the results of \cite{Lindstrom:2007vc} (as summarized and clarified in 
Appendix \ref{reduce}), we obtain the $\nll$ reduction of this action in the Wess-Zumino gauge; the part from $K^{(g)}$ is
\begin{eqnarray}
\label{eq::reduced_density}
\mathcal{L} \aleq \left( \Xi_+^A + \hcd{+}\hPh{i} E_{iC} E^{CA} \right) 
E_{AB} \left( \Xi_-^B + E^{BD} E_{Dj} \hcd{-}\hPh{j}  \right) \nn\\[1mm]
&& + \hcd{+}\hPh{i} \left( E_{ij} - E_{iA} E^{AB} E_{Bj}\right) \hcd{-} 
\hPh{j} \nn \\
&& + i K_i k^j \left( \dq{\phi} (\jj{+}{i}{j}+\jj{-}{i}{j}) +\dq{\chi} (\jj{+}{i}{j}-\jj{-}{i}{j}) + \dq{\prime} \Pi^i{}_j \right) 
\end{eqnarray}
where we introduce the matrices:
\begin{eqnarray}
E_{kl} \aleq K_{ij} \left( \jj{+}{i}{k}\jj{-}{j}{l} - \ft{1}{2} \Pi^i{}_k \delta^j{}_l - \ft{1}{2} \Pi^j{}_l \delta^i{}_k  \right) \\[2mm]
E_{Al} \aleq K_{ij}k^k \left( \begin{array}{c}
 \jj{-}{i}{k} \jj{-}{j}{l} \\[1mm]
 \Pi^i{}_k \jj{-}{j}{l}
\end{array}
\right) \\[2mm]
E_{kA} \aleq K_{ij}k^l \Big( \jj{+}{i}{k}\jj{+}{j}{l} ~,~ \jj{+}{i}{k} \Pi^j{}_l \Big) \\[2mm]
E_{AB} \aleq K_{ij}k^k k^l \left( \begin{array}{cc}
\jj{-}{i}{k} \jj{+}{j}{l} & \jj{-}{i}{k} \Pi^j{}_l \\[1mm]
\Pi^i{}_k \jj{+}{j}{l} & \Pi^i{}_k \Pi^j{}_l
\end{array} \right)
\end{eqnarray}
where the normalizations of the auxiliary fields $\Xi_\pm,\dq{}$ as well as the field-strength $f$
are given in Appendix \ref{reduce}.

The constraint reduces to
\begin{eqnarray}
 \mathcal{L}_{const.} \aleq 
 \tilde{X}_L (i\dq{\prime}-\ft{i}{2}f+\dq{\phi}-\dq{\chi} -iD_+\hcc{-})
 + \tilde{\psi}_- (+i \hee{+} - \hcc{+}) \nn \\
& +& \bar{\tilde{X}}_L (i\dq{\prime}-\ft{i}{2}f-\dq{\phi}+\dq{\chi} +iD_+\hcc{-})
+ \bar{\tilde{\psi}}_- (-i \hee{+} - \hcc{+}) \nn \\
& +& \tilde{X}_R (i\dq{\prime}+\ft{i}{2}f+\dq{\phi}+\dq{\chi} +iD_-\hcc{+})
+ \tilde{\psi}_+ (-i \hee{-} + \hcc{-}) \nn \\
& +& \bar{\tilde{X}}_R (i\dq{\prime}+\ft{i}{2}f-\dq{\phi}-\dq{\chi} -iD_-\hcc{+})
+ \bar{\tilde{\psi}}_+ (+i \hee{-} + \hcc{-}) ~,
\end{eqnarray}
where $\tilde X=\bbXh{}|$, $\tilde \psi_+=Q_+\bbXh{L}|$ and $\tilde \psi_-=Q_-\bbXh{R}|$
are the $\nll$ components of the Lagrange multipliers $\bbXh{}$.
\subsubsection{T-duality for the large vector multiplet in $\nll$ superspace} Integrating out the auxiliaries $\tilde{\psi}_\pm$ simply constrains $\Xi_{\pm}^A$ to vanish, and we obtain the gauged Lagrange density:
\begin{eqnarray}
\mathcal{L}\aleq K_{ij}(\jj{+}{i}{k} \jj{-}{j}{l}-\ft{1}{2} \Pi^i{}_k \delta^j{}_l
- \ft{1}{2} \delta^i{}_k \Pi_j{}^l ) \hcd{+}\hPh{k} \hcd{-}\hPh{l} \nn \\[1mm]
&+& i \dq{\phi}\big( K_i (\jj{+}{i}{j}+\jj{-}{i}{j}) k^j +i(\tilde{X}_L-\bar{\tilde{X}}_L+\tilde{X}_R-\bar{\tilde{X}}_R) \big) \nn\\[1mm]
&+& i\dq{\chi} \big(  K_i (\jj{+}{i}{j}-\jj{-}{i}{j}) k^j -i(\tilde{X}_L-\bar{\tilde{X}}_L-\tilde{X}_R+\bar{\tilde{X}}_R) \big) \nn\\[1mm]
&+& i\dq{\prime} \big(  K_i \Pi^i{}_j k^j 
-(\tilde{X}_L+\bar{\tilde{X}}_L+\tilde{X}_R+\bar{\tilde{X}}_R) \big) \nn\\[1mm]
&+& \ft{i}{2} f (\tilde{X}_L+\bar{\tilde{X}}_L-\tilde{X}_R-\bar{\tilde{X}}_R) \big)
%+ \dq{L}[iK_i\jj{-}{i}{j}k^j -(\tilde{X}_L-\bar{\tilde{X}}_L)] \nn \\[1mm]
%&&~ + \dq{R} [(iK_i\jj{+}{i}{j}k^j 
%-(\tilde{X}_R-\bar{\tilde{X}}_R)] 
%+ i\dq{\prime} [K_i \Pi^i{}_j k^j 
%- (\tilde{X}_L+\bar{\tilde{X}}_L+\tilde{X}_R+\bar{\tilde{X}}_R)] \nn \\[1mm]
%&&~ - \ft{1}{2}f [-i\tilde{X}_L-i\bar{\tilde{X}}_L
%+i\tilde{X}_R+i\bar{\tilde{X}}_R]
\end{eqnarray}

Imposing the equations of motion for $\dq{\alpha}$, which again just give diffeomorphisms, we obtain a gauged nonlinear $\sigma$-model with constrained field strength which proves that the dual geometries are indeed related by a Buscher duality.

\subsection{T-duality along semichiral isometries $k_{LR}$}
In the presence of semichiral superfields we can no longer decompose the action of the gauged isometry as in the BiLP case (\ref{kfcK}) and separate the rigid piece which acts on the \kah potential with $\mathcal{L}_k$. An extensive treatment of non BiLP geometries is left for \cite{LRRvUZ}. Making the notation of \cite{Lindstrom:2007vc} compatible with the previous section we redefine the complex potentials\footnote{See Appendix \ref{table} for full details} and reduce in the Wess-Zumino gauge:
\begin{equation}
\begin{array}{llll}
\vv{L}|=0 ~~,	 & (Q_+\vv{L})|=2\G_+	& (Q_-\vv{L})|= 0~~, 	& Q_+Q_- \vv{L} = -2i(\dt{2}-\dt{1})\\
\vv{R}|=0 ~~,	 & (Q_+\vv{R})|=0 	& (Q_-\vv{R})|= 2\G_-~~,&Q_+Q_- \vv{R} = -2i(\dt{2}+\dt{1})\\
\vv{\prime~}|=0~~,& (Q_+\vv{\prime~})|=0& (Q_-\vv{\prime~})|=0~~,& Q_+Q_- \vv{\prime~} = -2i\dt{3}~.
\end{array}
\end{equation}
The $\nll$ gauge field-strength $f=i(D_+\G_-+D_-\G_+)$ obeys the Bianchi identity
\begin{equation}
 \left. i (\ff-\ffb+\fft-\fftb)\right| = f
\end{equation}
(the $\nZZ$ field-strengths $\ff,\fft$ are given in Appendix \ref{table}).
Following \cite{Lindstrom:2007vc} we write the constrained Lagrange density
\begin{equation}
K_\mathbb{X} \left( \bbX{L}+\bbXB{L}+\vv{L} , \bbX{R}+\bbXB{R}+\vv{R} , i(\bbX{L}-\bbXB{L}+\bbX{R}-\bbXB{R})+\vv{\prime}\right) - \tilde{\phi} \vv{} - \bar{\tilde{\phi}} \vvb{} - \tilde{\chi} \vvt{} - \bar{\tilde{\chi}} \vvtb{}
\end{equation}
which reduces to $\nll$:
\begin{eqnarray}
\mathcal{L} =  E_{ij}\hcd{+}\bbX{}^i\hcd{-}\bbX{}^j\!\!\!\!\!\!\!\!\! && + \dt{1} 
[ (-i\pa{L} -i\pa{\bar{L}} + i \pa{R} + i\pa{\bar{R}})K 
+ 2(\tilde{\phi}-\bar{\tilde{\phi}}) ]\nn\\[1mm]
&& + \dt{2} [ (i\pa{L} + i\pa{\bar{L}} + i \pa{R} 
+ i\pa{\bar{R}})K - 2(\tilde{\chi}-\bar{\tilde{\chi}}) ]\nn\\[1mm]
&& + \dt{3} [ \ft{1}{2}(\pa{L} -\pa{\bar{L}} - \pa{R} 
+ \pa{\bar{R}})K 
-i (\tilde{\phi}+\bar{\tilde{\phi}}+\tilde{\chi}+\bar{\tilde{\chi}})] \nn\\[1mm]
&&+f[-i\hat{\phi}-i\bar{\tilde{\phi}}+i\tilde{\chi}+i\bar{\tilde{\chi}}]~,
\end{eqnarray}
where $E_{ij}=(g_{ij}+B_{ij})$ is the metric and 
$B$-field of the generalized \kah geometry as given 
in, \eg \cite{Lindstrom:2005zr}. As in the previous 
section, we impose the equations of motion for $\dt{\alpha}$ 
to obtain the gauged nonlinear $\sigma$-model with 
the constraint on the field-strength $f$ that we 
recognize as the hallmark of T-duality. Again, 
the $\dt{\alpha}$ equations of motion just give diffeomorphisms. 

\section{Conclusions}
We have used the gauge multiplets constructed in \cite{Lindstrom:2007vc,Gates:2007ve} 
to investigate the duality between semichiral and (twisted) chiral superfields 
discovered in \cite{Grisaru:1997ep}, and found that the dual geometries are 
related by Buscher duality. We demonstrated this in $\nZZ$ superspace where we gave
the generalized \kah potentials with gauged isometries. 
When we descended to $\nll$ superspace, the nature of the T-duality was 
clarified: we found a gauged nonlinear $\sigma$-model with a Lagrange 
multiplier constraining the field-strength of the gauge field as well as 
diffeomorphisms relating the generalized moment maps in the original 
geometry to natural coordinates in the dual geometry.

This work is part of an ongoing exploration of generalized complex geometry,
using nonlinear $\sigma$ models, and is therefore complimentary 
to the mathematical aspects of T-duality considered in \cite{cavalcanti}. 
The full construction of the moment maps and a geometric discussion 
of these results is left for future work \cite{LRRvUZ}.

\bigskip\bigskip
\noindent{\bf\large Note}:

\bigskip
\noindent 
After completing our work, we became aware of
related results obtained by W.~Merrell and D.~Vaman.

\bigskip\bigskip
\noindent{\bf\Large Acknowledgement}:
\bigskip\bigskip

\noindent  
UL supported by EU grant (Superstring theory)
MRTN-2004-512194 and VR grant 621-2006-3365.
The work of MR and IR was supported in part by NSF grant no.~PHY-0354776. 
The research of R.v.U. was supported by 
Czech ministry of education contract No.~MSM0021622409. 
The research of M.Z. was
supported by VR-grant 621-2004-3177. 

\appendix
\Section{Reduction to $\nll$ superspace for gauged BiLP geometries}
\label{reduce}
In this appendix we review some of the results of \cite{Lindstrom:2007vc} as they emerge from inherent geometric objects for BiLP geometries. The $\nll$ invariants system of \cite{Lindstrom:2007vc} is slightly modified so that the reduction of the gauged Lagrange density (\ref{eq:large_kahler_pot}) to $\nll$ is simpler in this context; namely, carrying out the reduction for the matter couplings piece will give convenient redefinitions for $\nll$ gauge invariants. Acting with $Q_\pm$ on the gauged action we can identify the connections $A_\pm$ that enter with $J_\pm k$ respectively:
\begin{eqnarray}
Q_\pm K^{(g)} = \!\!\!&&\!\!\!\!\!\! K^{(g)}_i \left( \jj{\pm}{i}{j} 
D_\pm \varphi^j -\ft{1}{4} Q_\pm(V^\phi+V^\chi)\jj{+}{i}{j}k^j \right. \nn \\ 
&&  ~~\left. -\ft{1}{4}Q_\pm(V^\phi-V^\chi)\jj{-}{i}{j}k^j - 
\ft{1}{4}Q_\pm V^\prime \Pi^i{}_j k^j  \right) \nn \\[1mm] 
= \!\!\!&&\!\!\!\!\!\! K^{(g)}_i \left( \jj{\pm}{i}{j} 
\hcd{\pm} \varphi^j + \hee{\pm} \jj{\mp}{i}{j}k^j + \hcc{\pm} \Pi^i{}_j k^j \right) ~.
 \end{eqnarray}
We find it useful to modify the $\nll$ notation of \cite{Lindstrom:2007vc}, introducing:
\begin{equation}
\label{eq::q_invariants}
\dq{\phi} = -i\ft{1}{2} (Q_{[+}\hee{-]}-D_{[+}\hcc{-]}) ~~,~~~ \dq{\chi} = -i\ft{1}{2}(Q_{(-}\hee{+)}+D_{(+}\hcc{-)}) ~~,~~~\dq{\prime} = -i\ft{1}{2} Q_{[+}\hcc{-]} \end{equation}
%
%\begin{eqnarray}
%\dq{\phi} \aleq \ft{1}{2}(-iQ_+\hee{-}-iD_-\hcc{+}+iQ_-\hee{+}+iD_+\hcc{-}) \nn \\
%\dq{\phi} \aleq -i\ft{1}{2} (Q_{[+}\hee{-]}-D_{[+}\hcc{-]}) \nn \\
%\dq{\phi} \aleq \ft{i}{2}(Q_{[-}\hee{+]}+D_{[+}\hcc{-]}) \nn \\
%\dq{\chi} \aleq \ft{1}{2}(-iQ_+\hee{-}-iD_-\hcc{+}-iQ_-\hee{+}-iD_+\hcc{-}) \nn \\
%\dq{\chi} \aleq -i\ft{1}{2}(Q_{(-}\hee{+)}+D_{(+}\hcc{-)}) \nn \\
%\dq{\chi} \aleq \ft{-i}{2}(Q_{[-}\hee{+]}+D_{[+}\hcc{-]}) \nn \\
%\dq{\phi} \aleq \ft{i}{2}(Q_{[-}\hee{+]}+D_{[+}\hcc{-]}) \nn \\
%\dq{\prime} \aleq -i\ft{1}{2} Q_{[+}\hcc{-]}
%\end{eqnarray}
%
% \begin{equation}
% \dq{L} \equiv i Q_- \hee{+}   ~~,~~~\dq{R} \equiv -i Q_+ \hee{-}~~,~~~ \dq{\prime} = -\ft{i}{2} %(Q_+ \hcc{-} - Q_- \hcc{+}) ~,
%\end{equation}
and the field-strength for the connections $A_\pm$
\begin{equation}
 f=-iQ_{(+} \hcc{-)} = i (D_+ A_- + D_- A_+)
\end{equation}
which allows us to write the reduction for $Q_+Q_- K^{(g)}$ in terms of the geometric objects:
\begin{eqnarray}
&& Q_+Q_- K^{(g)} = \nn \\
&& \quad K_{ij}^{(g)} \big[ (\jj{+}{i}{k} \hcd{+}\hPh{k}+\jj{-}{i}{k}k^k \hee{+}+\Pi^i{}_k k^k \hcc{+}) (\jj{-}{j}{l} \hcd{-}\hPh{k}+\jj{+}{j}{l}k^l \hee{-}+\Pi^j{}_l k^l \hcc{-}) \nn \\
&& \quad \quad \quad ~  -\ft{1}{2}(\delta^i{}_k\Pi^j{}_l+\Pi^i{}_k \delta^j{}_l)\hcd{+}\hPh{k} \hcd{-}\hPh{l} \big]  \nn \\
&& \quad + iK^{(g)}_i k^k \big( \dq{\phi} (J_+ + J_-)^i{}_k+\dq{\chi} (J_+ - J_-)^i{}_k+\dq{\prime} \Pi^i{}_k \big)
\end{eqnarray}

%\begin{eqnarray}
% Q_+ Q_- \varphi^i \aleq \ft{1}{2} \Pi^i{}_j D_+ ( \hcd{-} \varphi^j -\hee{-}k^j - \hcc{-} %\jj{-}{i}{j}k^j ) - (+\leftrightarrow-) \nn \\
% && +i\dq{R} \jj{+}{i}{j}k^j + i \dq{L}\jj{-}{i}{j}k^j + i\dq{\prime} \Pi^i{}_j k^j
% \end{eqnarray}
\Section{Conventions and notation}
\label{table}
The conversion between the notation of \cite{Lindstrom:2007vc} and the current notation can
be derived from changing some signs:
\begin{equation}
\left\lbrace  \vv{\prime} , \vv{R} , V^\prime , V^R \right\rbrace \rightarrow - \left\lbrace  \vv{\prime} , \vv{R} , V^\prime , V^R \right\rbrace  
\end{equation}
as well as
\begin{equation}
\{ \lamt , \lam^R \} \rightarrow - \{ \lamt , \lam^R \}~.
\end{equation}
These changes correct some unnatural conventions for the definitions of isometries.

We summarize the essential consequences here for both the large vector multiplet and the semichiral vector multiplet in the tables below.

 \begin{table}
 \begin{center}
 \begin{tabular}{|c||c|c|}
 \hline
Object			&Old			
 				&New \\ \hline \hline &\multicolumn{2}{|c|}{}\\
$\delta V^\phi$		&\multicolumn{2}{|c|}{$i(\lamb-\lam)$}\\[2mm]
$\delta V^\chi$		&\multicolumn{2}{|c|}{$i(\lamtb-\lamt)$}\\[2mm] \hline &&\\
$\delta V^\prime$	&$\lam+\lamb+\lamt+\lamtb$
 				& $-\lam-\lamb+\lamt+\lamtb$\\[2mm] \hline &&\\
Complex potential 	&$ V = \ft{1}{2} (V^\prime + i (V^\phi+V^\chi))$
				&$ V_L=\ft{1}{2} (-V^\prime + i (V^\phi-V^\chi))$ \\[2mm]
and variation (1)	&$\delta V = \lam+\lamt$
				&$\delta V_L = \lam-\lamt$\\[2mm] \hline &&\\ 
Complex potential	&$ \tilde{V} = \ft{1}{2} (V^\prime + i (V^\phi-V^\chi))$
				&$ V_R=\ft{1}{2} (-V^\prime + i (V^\phi+V^\chi))$ \\[2mm] 
and variation (2)	&$\delta V = \lam+\lamtb$
				&$\delta V_L = \lam-\lamtb$\\[2mm] \hline &&\\
 $\nZZ$			&$ \bbG{+} = \bbDB{+} V $ 
 				&$ \bbG{+} = \bbDB{+} V_L $\\[2mm]
Gauge invariants	&$ \bbG{-} = \bbDB{-} \tilde{V} $ 
				&$ \bbG{-} = \bbDB{-} V_R $\\[2mm]
 			&$ \bbGB{+} = \bbD{+} \bar{V} $ 
				& $ \bbGB{+} = \bbD{+} \bar{V}_L $\\[2mm]
			& $ \bbGB{-} = \bbD{-} \bar{\tilde{V}} $ 
				& $ \bbGB{-} = \bbD{-} \bar{V}_R $\\[2mm]\hline &\multicolumn{2}{|c|}{}\\
Decomposition 		& \multicolumn{2}{|c|}{$\Xi^A_{\pm} = \left(\,\left. \hbox{Re}
			(\bbG{\pm})\right| ,\left. \hbox{Im} (\bbG{\pm}) \right|\,\right)$} \\[2mm]
to $\nll$ 		& \multicolumn{2}{|c|}{$D_\pm \Xi_\mp^{A}$} \\[2mm]\hline &&\\
			& $\dq{1} = i(Q_- \hee{+} - Q_+ \hee{-}) $
			& $\dq{\phi} = -i\ft{1}{2} (Q_{[+}\hee{-]}-D_{[+}\hcc{-]}) $ \\[2mm] 
$q$-invariants:		& $\dq{2} = i(Q_- \hee{+} + Q_+ \hee{-}) $
				& $\dq{\chi} = -i\ft{1}{2}(Q_{(-}\hee{+)}+D_{(+}\hcc{-)})$ \\[2mm] 
			& $\dq{3} = i(Q_- \hcc{+} - Q_+ \hcc{-}) $
				& $\dq{\prime} = -i\ft{1}{2}(Q_+\hcc{-}-Q_-\hcc{+}) $\\[2mm] \hline &&\\
The field-strength $f$ 	&$i(Q_+ \hcc{-} + Q_- \hcc{+} )$ 
				& $-i(Q_+ \hcc{-}+Q_- \hcc{+})$ \\[2mm] \hline
\end{tabular}
\end{center}
\caption{Large vector multiplet conventions and definitions}
\end{table}
\begin{table}
\begin{center}
\begin{tabular}{|c||c|c|}
 \hline
Object&Old&New\\ \hline \hline &\multicolumn{2}{|c|}{}\\
$\delta \vv{L}$&\multicolumn{2}{|c|}{$i(\lamb_L-\lam_L)$}\\[2mm]
$\delta \vv{R}$&\multicolumn{2}{|c|}{$i(\lamb_R-\lam_R)$}\\[2mm] \hline &&\\
$\delta \vv{\prime}$ & $\lam_L+\lamb_L+\lam_R+\lamb_R$ & $ -\lam_L -\lamb_L + \lam_R + \lamb_R$\\[2mm] \hline &&\\
Complex potential 	&$ \vv{} = \ft{1}{2} (\vv{\prime} + i (\vv{L}+\vv{R}))$
			&$ \vv{} = \ft{1}{2} (-\vv{\prime} + i (\vv{L}-\vv{R}))$\\[2mm]
and variation (1)	&$\delta \vv{} = \lam_L+\lam_R$
				&$\delta \vv{} = \lam_L-\lam_R$\\[2mm] \hline &&\\
Complex potential 	&$ \vvt{} = \ft{1}{2} (\vv{\prime} + i (\vv{L}-\vv{R}))$
			&$ \vvt{} = \ft{1}{2} (-\vv{\prime} + i (\vv{L}+\vv{R}))$\\[2mm]
and variation (2)	&$\delta \vv{} = \lam_L+\lamb_R$ &$\delta \vv{} = \lam_L-\lamb_R$\\[2mm] \hline&\multicolumn{2}{|c|}{}\\
 $\nZZ$			& \multicolumn{2}{|c|}{$\ff = \bbDB{+}\bbDB{-}\vv{}$ ~~,~~~ $\ffb = 
 -\bbD{+}\bbD{-}\vvb{}$}\\[2mm]
Gauge invariants	&\multicolumn{2}{|c|}{$\fft = \bbDB{+}\bbD{-} \vvt{}$ ~~,~~~ $\fftb = - \bbD{+}\bbDB{-}\vvtb{}$}\\[2mm]\hline &\multicolumn{2}{|c|}{}\\
$\dt{}$-invariants & \multicolumn{2}{|c|}{$\dt{1}=\!\left.\left(\ff+\ffb\right)\right| ~~,~~~ \dt2=\!\left.\left(\fft+\fftb\right)\right| ~~,~~~ \dt3=\!\left.i\!\left(\ff-\ffb-\fft+\fftb\right)\right|$}\\[4mm] \hline &&\\
Gauge fields & $\G_+=\ft{1}{2}\left.\left(Q_+\vv{L}-\ft{1}{2}D_+\vv{\prime}\right)\right|$& $\G_+=\ft{1}{2}\left.\left(Q_+\vv{L}+\ft{1}{2}D_+\vv{\prime}\right)\right|$ \\[2mm]
~&$\G_-=-\ft{1}{2}\left.\left(Q_-\vv{R}-\ft{1}{2}D_-\vv{\prime}\right)\right|$ &$\G_-=\ft{1}{2}\left.\left(Q_-\vv{R}-\ft{1}{2}D_-\vv{\prime}\right)\right| $\\[2mm] \hline &\multicolumn{2}{|c|}{}\\
Bianchi identity & \multicolumn{2}{|c|}{$\left. i(\ff-\ffb+\fft-\fftb)\right| = f = i (D_+ \G_- + D_- \G_+)$}\\[6mm] \hline
\end{tabular}
\end{center}
\caption{Semichiral vector multiplet conventions and definitions}
\end{table}

\end{document}